\documentclass[english]{article}
\usepackage[T1]{fontenc}
\usepackage[latin9]{inputenc}
\usepackage[a4paper]{geometry}
\usepackage{color,ulem}
\usepackage{array}
\usepackage{amsbsy}
\usepackage{amstext}
\usepackage{stmaryrd}
\usepackage{stackrel}
\usepackage{graphicx}

\makeatletter

\newcommand{\lyxmathsym}[1]{\ifmmode\begingroup\def\b@ld{bold}
  \text{\ifx\math@version\b@ld\bfseries\fi#1}\endgroup\else#1\fi}

\providecommand{\tabularnewline}{\\}

\@ifundefined{date}{}{\date{}}

\usepackage{cite,color}
\usepackage{times}

\usepackage{babel}

\@ifundefined{showcaptionsetup}{}{%
 \PassOptionsToPackage{caption=false}{subfig}}
\usepackage{subfig}
\makeatother

\usepackage{babel}
\begin{document}
	
\title{Noisy three-player dilemma game: Robustness of the quantum advantage \color{black}} 
\author{{\normalsize{}{}Pranav Kairon$^{\mathsection}$, Kishore Thapliyal$^{\ddagger,}$\thanks{Email: kishore.thapliyal@upol.cz}},{\normalsize{}{} R. Srikanth$^{\natural}$,
and Anirban Pathak$^{\spadesuit,}$}\thanks{Email: anirban.pathak@jiit.ac.in}{\normalsize{}{}
}\\
 {\normalsize{}{}$^{\mathsection}$Delhi Technological University, Bawana Road, Delhi, 110042, India}\\
 {\normalsize{}{}$^{\ddagger}$RCPTM, Joint Laboratory of Optics
of Palacky University and Institute of Physics of Academy}\\
 {\normalsize{}{}of Science of the Czech Republic, Faculty of Science,
Palacky University, }\\
 {\normalsize{}{}17. listopadu 12, 771 46 Olomouc, Czech Republic}\\
 {\normalsize{}{}$^{\natural}$Poornaprajna Institute of Scientific Research, Bengaluru, Karnataka 560080, India}\\
 {\normalsize{}{}$^{\spadesuit}$Jaypee Institute of Information
Technology, A-10, Sector-62, Noida, UP-201309, India}}
\maketitle
\begin{abstract}
{\normalsize{}{}Games involving quantum strategies often yield higher
payoff. Here, we study a practical realization of the three-player dilemma game using the superconductivity-based quantum processors provided by IBM Q Experience. We analyze the persistence of the quantum advantage under corruption of the input states and how this depends on parameters of the payoff table. Specifically, experimental fidelity and error are observed not to be properly anti-correlated, i.e., there are instances where a class of experiments with higher fidelity yields a greater error in the payoff. Further, we find that the classical strategy will always outperform the
quantum strategy if corruption is higher than half. }

\end{abstract}

\section{{\large{}{}Introduction}}

Game theory provides a way to learn about decisive communication between
rational and self-seeking agents. Therefore, it plays an important
role in the fields of computer science, economics, biology, psychology,
etc. (see \cite{flitney2002introduction,piotrowski2003invitation}
for review). Computationally, game theory can be used to model algorithms
\cite{li2012application,elhenawy2015intersection} as well as check the robustness
of networks and corresponding attack strategies \cite{laszka2012game}.
In cryptography, the communication task can be visualized as a game between the
parties trying to communicate securely and an eavesdropper (\cite{dodis2007cryptography}
and references therein).  With the advent of quantum computing, it is observed that resources used in quantum computing,
such as quantum coherence and entanglement, provide alternative solutions
to classical games. 

We may mention, for example, the emergence of cooperation in the prisoner's
dilemma game \cite{li2014entanglement} and the resolution of the coordination
in battle of sexes game \cite{nawaz2004dilemma} using entanglement. 
Specifically, as
all the players wish to maximize their gain or payoff in games, for
which the umpire has laid down the rule(s),  players using quantum mechanical
tactics are found to attain a higher payoff compared to the classical one \cite{meyer1999quantum}. 
Further, the dilemma disappears in prisoner's dilemma with the use
of quantum resources under unitary operations  \cite{eisert1999quantum,du2002playing}.
Along the same line, optimal cloning of quantum states is also studied
as game \cite{werner1998optimal}. Quantum games based on monogamy
of entanglement are shown to be useful in device independent quantum cryptography
\cite{tomamichel2013monogamy}. Our understanding of several other foundational aspects of quantum mechanics is improved by considering games, such as nonlocality \cite{fritz2013local}, {the uncertainty bound on nonlocality \cite{oppenheim2010uncertainty},} contextuality \cite{anshu2020contextuality}, PR-boxes \cite{popescu2014nonlocality}, as well as applications in quantum reinforcement learning \cite{chen2006quantum} and quantum machine learning \cite{clausen2018quantum}.

Over the course of time, multiplayer quantum games were also introduced {that exploit quantum correlation to prevent betrayal by individual players
\cite{benjamin2001multiplayer}. It has been suggested that these quantum
games may shed light} on the interactions in many-particle systems \cite{johnson2001playing}. One such multiplayer
game is the three-party counterpart of the prisoner's dilemma. In the classical version, all three players prefer to choose analogous to the corresponding two-party case. The dilemma exists
because the Nash equilibrium does not coincide with the
Pareto optimal \cite{benjamin2001multiplayer}. Specifically, a Nash equilibrium is the situation in which no
participant can gain by a unilateral change of strategy, while Pareto optimal corresponds to the situation that any change
in strategy would make at least one individual worse off \cite{benjamin2001multiplayer}. 
Still in quantum case, use of tripartite entanglement shows certain
advantage. Moreover, computing the Nash equilibrium in the three- and four- player
games is shown to be a hard problem \cite{daskalakis2005three,daskalakis2009complexity}.
An experimental verification of three-player dilemma game using NMR was
 reported in Ref. \cite{mitra2007experimental}.  In the recent past,
other games have been realized on photonic quantum computer \cite{kolenderski2012aharon,pinheiro2013vector,schmid2010experimental,zhou2003proposal}
and ion trap platform \cite{solmeyer2018demonstration}.

In general, the dilemma games are relevant in several studies of biology,
economics, psychology, international relations, sports to name a few.
For instance, King Solomon's dilemma \cite{glazer1989efficient} based
on the Old Testament can model prize allocation, research grant distribution,
etc. Another multiparty version of prisoner's dilemma is diner's dilemma
in which each player has to choose whether to order an expensive or
an inexpensive dish if they have to equally share the bill \cite{teng2013trust}.
This iterated diner's dilemma is useful in the social dynamics of networks
and situational awareness. Such iterated multiparty prisoner's dilemma
in the context of social dynamics is discussed in the past, too \cite{bankes1994exploring}.
Along the same line, dilemma of the players in other games is used
to introduce the conditional probability \cite{morgan1991let}.

Decoherence is the Achilles' heel of quantum computing and information
processing in particular, and technology in general. Similar results
are shown for the quantum games \cite{ozdemir2004quantum}. Independently,
the effect of errors in the initial state preparation (as corruption
by a demon) on the outcome of three-player dilemma game is studied assuming
that the players are unaware of corruption and that there is no decoherence
\cite{johnson2001playing}. Interestingly, beyond a pivotal value
of corruption it can be observed that players fare off better with
the classical strategies, but since players have no knowledge of the  level
of corruption they have to stick to their original strategies. Furthermore,
a quantum game reduces to classical game if one of the parties
allows his qubit to decohere under Markovian noise channels \cite{chen2002noisy},
while Nash equilibria are unchanged by decoherence for prisoner's
dilemma \cite{chen2003quantum}. 

Here, we wish to implement the three-player dilemma game \cite{benjamin2001multiplayer}
on IBM quantum computer \cite{IBMQ} and study how the change in the utility
function affects the point of quantum advantage. Interestingly, this is the first
realization of a game with corrupt source on a superconducting qubits
based quantum computer. Despite high error rate and the limited qubit
connectivity, it has been shown to run a wide array of algorithms
(\cite{sisodia2017design,sisodia2018circuit} and references therein). Thus, we realize the game on IBM Q Experience and compare
the experimental payoffs with previous experiments on NMR \cite{mitra2007experimental}.
On generalizing the payoff table in the noisy game, the point where
quantum advantage disappears also changes which leads to some interesting
observations. 
{In specific, they show how robust the quantum strategy is. An application of these results is that given a known corruption level, the payoff table (the relative stakes) may, in a range, be chosen to give an advantage to the quantum strategy.}
Finally, we show that classical strategies
dominate when corruption is higher than 50\% in the proposed game.

The rest of the paper is organized as follows. We introduce three-player
quantum dilemma in Section \ref{sec:3-Person-Quantum-Dilemma}. The noisy
counterpart of the game and its experimental implementation is discussed
in Section \ref{sec:CORRUPTION-AND-IBM}. We further discuss all the
results in detail in the penultimate section before concluding the paper
in Section \ref{sec:Conclusion-and-Discussion}.

\section{{\large{}{}Three-person quantum dilemma \label{sec:3-Person-Quantum-Dilemma}}}

A multiparty dilemma game was introduced as a multiparty counterpart of the prisoner's
dilemma game, where each person has two choices: either
to cooperate (0) or defect (1). The three-player dilemma resembles el-Farol
bar problem that players have to decide independently whether to go
or not to a bar with seating capacity for only two (\cite{mitra2007experimental}
and references therein).

In the three-player quantum dilemma game, each player is provided one qubit
by the umpire, who
performs an entangling operation on state $\left|000\right\rangle $ before that which increases the nonclassical correlation
among the players. The entangling gate $J$ can be defined as in Ref. \cite{benjamin2001multiplayer}:
\begin{equation}
J=\cos\frac{\gamma}{2}I^{\varotimes3}+i\sin\frac{\gamma}{2}X^{\varotimes3},\label{eq:J}
\end{equation}
where $I$ and $X$ are identity and Pauli NOT gates, respectively.
Without loss of generality, we choose the case when the correlation
is maximum, i.e., $\gamma=\pi/2$. Further, it can be checked that
for minimum correlation, i.e., $\gamma=0$, the game reduces to its
classical counterpart
\cite{du2002playing}. 

\begin{figure}
\begin{centering}
\includegraphics[scale=0.5]{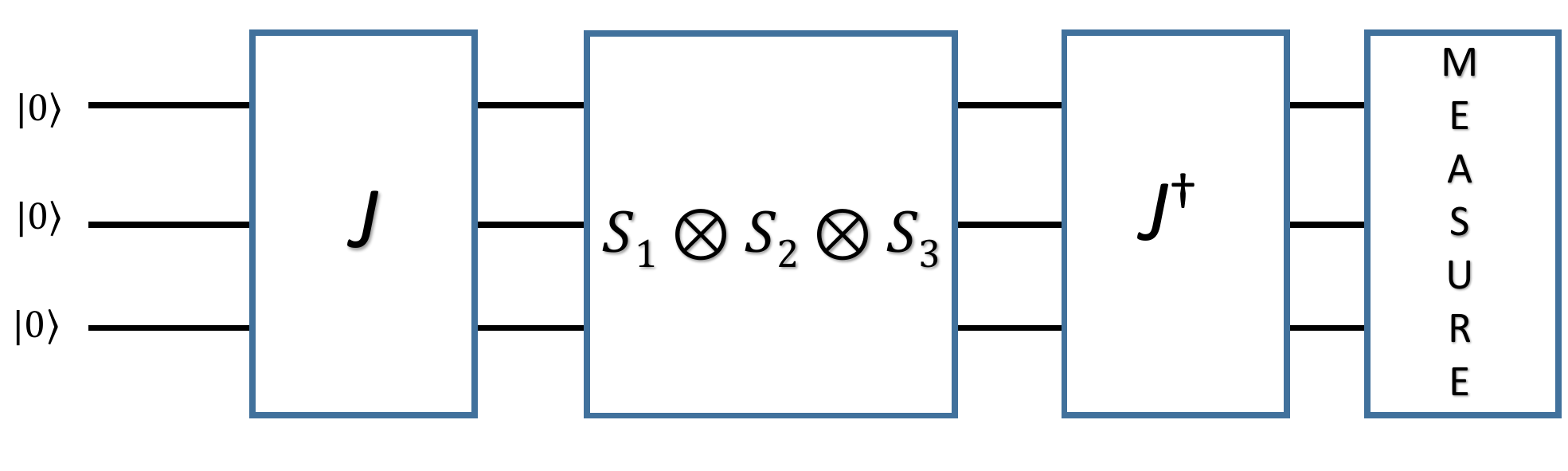} 
\par\end{centering}
\caption{\label{fig:circuit}(Color online) Circuit diagram that can be used for the realization of the three-player
quantum dilemma game.}
\end{figure}

In quantum game, each person is allowed to choose an operation from
a strategy set $S$, consisting of 3 elements $S=(S_{1},S_{2},S_{3})$,
where $S_{1}=X$ means the player wants to attend the party; $S_{2}=H$
corresponds to the player's choice to go with half a probability;
and $S_{3}=I$ represents the player wants to stay at home. Note that
the choice of $S_{2}$ does not have a counterpart in classical games.
This is a restricted strategy set (as there can be an infinitely many
possible quantum strategies each corresponding to a different unitary
operation), but it encompasses all the {nonclassical} characteristics we want to demonstrate
through this game. Subsequently, a disentangling operation $J^{\dagger}=\frac{I^{\varotimes3}-iX^{\varotimes3}}{\sqrt{2}}$
is performed before measuring in the computational basis. The circuit
diagram of the game is shown in Fig. \ref{fig:circuit}.

Thus, when none of the players decide to go, i.e., the measurement
outcome is $\left|000\right\rangle $ (represented by corresponding
bit values 000 in Table \ref{tab:Generalized-Payoff-Table}), nobody
is happy since they could not attend the party but are not sad
since none of the friends betrayed, and thus everybody gets $0$ payoff.
However, if one person decides to go then the other two will be unhappy
(with payoff $-n$), and the one attending the party does not enjoy
being alone (with payoff $p$). When two of the friends decide to
go they both fare off with $n$ payoff each since they get to go to
the party with company, while the friend left behind is not too dejected
since his presence would have overcrowded the party so he gets $p$.
If all of them decide to go, they get payoff of $q$ each since their
presence has overcrowded the party. {Accordingly, we impose $n > q > p >0$. } This problem is also relevant in the recent pandemic coronavirus situation that may allow a restaurant to open but to avoid infection it restricts people who can sit at a table to two, say because the table is 1 m wide and only two persons can sit in the same table opposite to each other without violating the social distancing norms. 

Yet another example for the
three-player dilemma closer to most of us would be the dilemma of three academic
collaborators in applying for a research grant. If two of them apply, they are likely to receive
the grant, whereas they probably would receive insufficient
or no funding if all three apply for it. Also, they would not be happy if
none of them apply or their collaborator gets it but not them.
The dilemma shown previously was by considering $n=9,\,p=1,$ and $q=2$
\cite{benjamin2001multiplayer}. Here, we study a general
description of such payoff tables and show how the payoff depends
on these parameters in the noisy game (subjected to constraint $0<p<q<n$).
{One motivation for this is to understand whether and how  the game's stakes can be fixed based on knowledge of the
preparation noise in the system.}
In \cite{benjamin2001multiplayer}, it is shown that in a special
case considering a different set of values of the payoffs, 
quantum players do not have any advantage over the classical strategy which is no longer a Nash equilibrium. Therefore, we have restricted ourselves
to the aforementioned constraint which ensures that classical Nash
equilibrium exists. To the best of our knowledge, this is the first
attempt to generalize the payoff table for the three-player dilemma game in analogy of prisoner's
dilemma \cite{tian2015monkeys}. 

{\small{}{}}{\small\par}

\begin{table}
\noindent \begin{centering}
{\small{}{}}%
\begin{tabular}{|>{\centering}p{5cm}|c|}
\hline 
{\small{}{}Bit values corresponding to possible measurement outcome}  & {\small{}{}Payoffs $\left(\$\right)$}\tabularnewline
\hline 
\hline 
{\small{}{}000}  & {\small{}{}$0,0,0$}\tabularnewline
\hline 
{\small{}{}001}  & {\small{}{}$-n,-n,p$}\tabularnewline
\hline 
{\small{}{}010}  & {\small{}{}$-n,p,-n$}\tabularnewline
\hline 
{\small{}{}011}  & {\small{}{}$p,n,n$}\tabularnewline
\hline 
{\small{}{}100}  & {\small{}{}$p,-n,-n$}\tabularnewline
\hline 
{\small{}{}101}  & {\small{}{}$n,p,n$}\tabularnewline
\hline 
{\small{}{}110}  & {\small{}{}$n,n,p$}\tabularnewline
\hline 
{\small{}{}111}  & {\small{}{}$q,q,q$}\tabularnewline
\hline 
\end{tabular}
\par\end{centering}
\begin{centering}
 
\par\end{centering}
{\small{}{}\caption{\label{tab:Generalized-Payoff-Table}Generalized payoff table for
three-player dilemma game depending upon the possible measurement outcomes.
In previous adaptations of the game \cite{benjamin2001multiplayer},
$n=9,\,p=1,$ and $q=2$ were used. }
}{\small\par}

\end{table}

\section{{\large{}{}IBM implementation {and noisy state preparation} \label{sec:CORRUPTION-AND-IBM}}}

\begin{figure}
\begin{centering}
\includegraphics[scale=0.5]{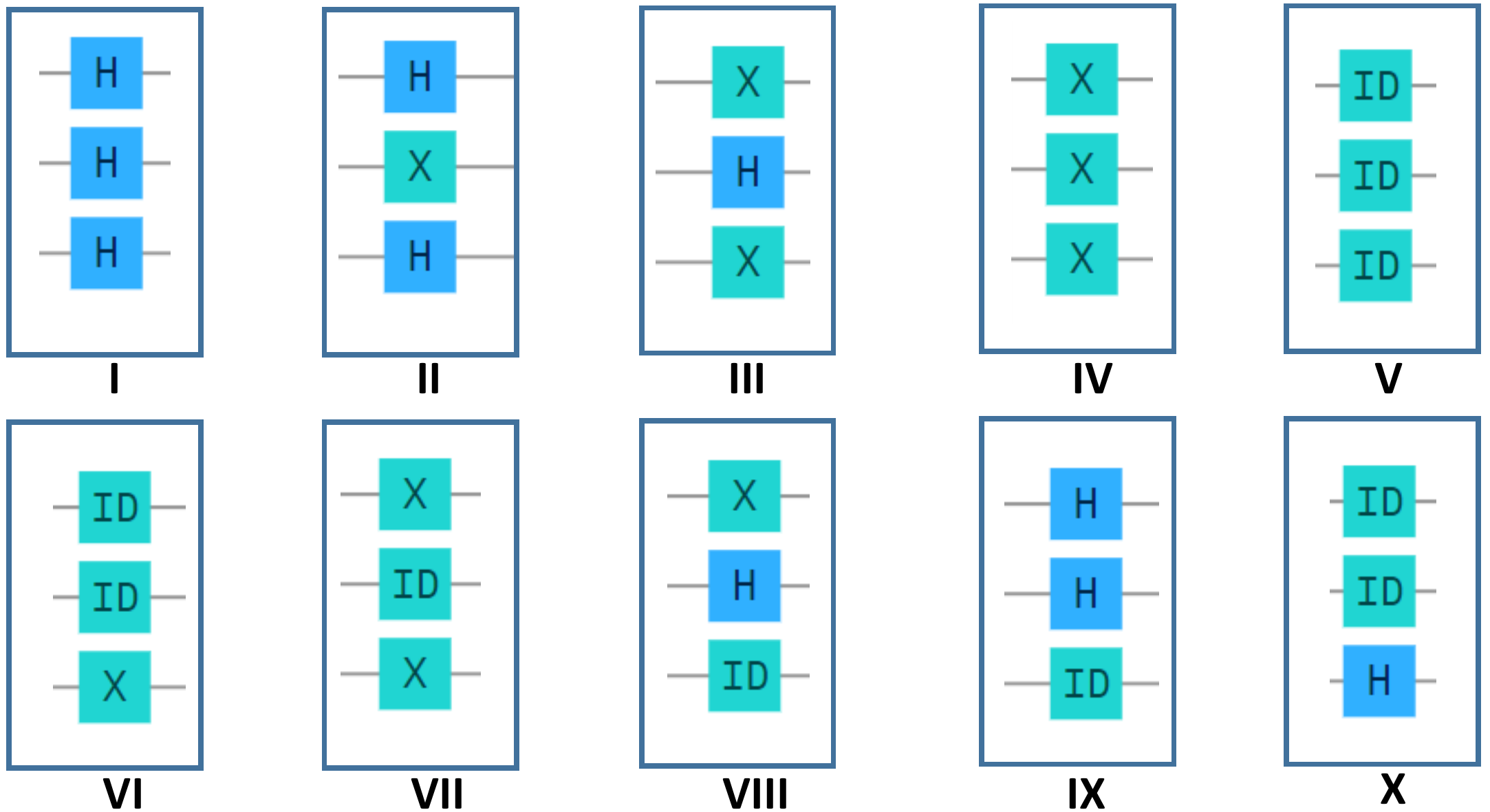} 
\par\end{centering}
\caption{\label{fig:10cl}(Color online) Ten classes with different operations
used by each player in $S_{1}\otimes S_{2}\otimes S_{3}$. Here, $H$,
$X$, and ${\rm ID}$ are Hadamard, NOT, and identity gates, respectively. }
\end{figure}

For the given strategy space $S$, there are 3 choices per player which
gives us $3^{3}=27$ arrangements which can be clustered into 10 different
classes \cite{johnson2001playing}. Experimental design for the implementation
of all these classes is shown in Fig. \ref{fig:10cl}. Classes
I, IV, and V have {~${3 \choose 3}=1$~} configuration, while Classes II, III,
VI, VII, IX and X each have { ${3 \choose 1} = 3$}  possible configurations, and there
are { $3! = 6$}  configurations in Class VIII. When each player decides to play
a unique tactic, independent of the other players, we obtain class
VIII as mixed strategy Nash Equilibrium. Class VII provides best
response for each player but is not considered as it is biased. Suppose
the umpire has provided tainted qubits from the black box, i.e., instead
of $\left|000\right\rangle $ he introduces error (mixedness) of the
form $(1-x)\left|000\right\rangle \left\langle 000\right|+x\left|111\right\rangle \left\langle 111\right|$.
The expected payoff would change quite exorbitantly as we increase the
amount of {preparation noise or }corruption (as shown in Fig. \ref{fig:x-c}), where $\lyxmathsym{\textquoteleft}x$'
is the noise parameter, $x\in[0,1]$.

\begin{figure}
\begin{centering}
\includegraphics[scale=0.8]{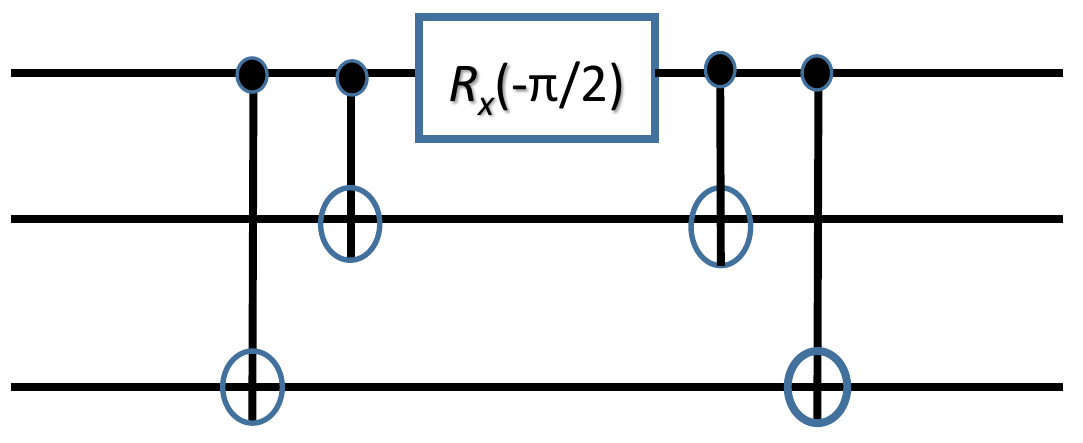} 
\par\end{centering}
\caption{\label{fig:J}(Color online) Decomposition of three qubit operation
$J$.}
\end{figure}
In IBM implementation \cite{shukla2018complete} of the gate \cite{sousa2006universal},
we performed the entangling gate $J$ using a $R_{x}(-\frac{\pi}{2})$
and 4 CNOT gates as 
\[
J=\left(I_{0}\otimes CNOT_{1\rightarrow2}\right)\,\left(CNOT_{1\rightarrow0}\otimes I_{2}\right)\,\left(I_{0}\otimes R_{x}\left(-\frac{\pi}{2}\right)_{1}\otimes I_{2}\right)\,\left(I_{0}\otimes CNOT_{1\rightarrow2}\right)\,\left(CNOT_{1\rightarrow0}\otimes I_{2}\right)
\]
on qubits 0, 1, and 2, respectively (also shown in the Fig. \ref{fig:J}).
The single qubit operation can be defined as $R_{x}(-\frac{\pi}{2})\left|0\right\rangle \rightarrow\left(\left|0\right\rangle +i\left|1\right\rangle \right)/\sqrt{2}$.
In fact, many simulations of this game have modeled the entangling
gate $J$ incorrectly in the past \cite{sousa2006universal} as those
matrix decomposition were not the same as $J$. As already discussed this is followed
by the players applying their operations on their respective qubits
from the strategy space $S$. Finally, we apply a disentagling gate
$J^{\dagger}$ which can be modeled by the same gates as $J$ since
$(ABCDE)^{\dagger}=E^{\dagger}D^{\dagger}C^{\dagger}B^{\dagger}A^{\dagger}$.

\begin{figure}
\begin{centering}
\includegraphics[scale=0.6]{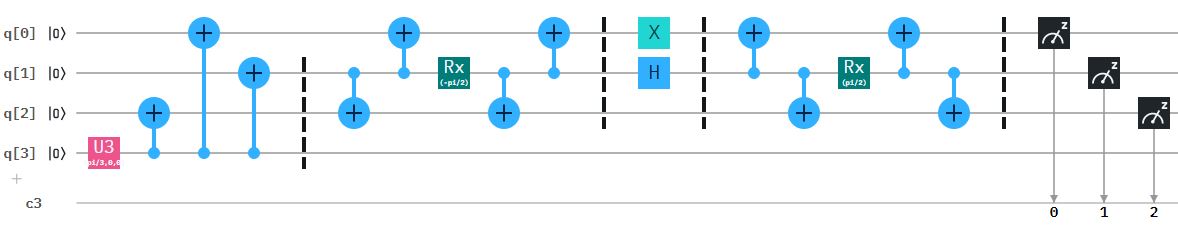} 
\par\end{centering}
\caption{\label{fig:corrupt}(Color online) Circuits for the simulation of
noisy three-player quantum dilemma game for Class VIII as implemented
on IBM Quantum Experience. }
\end{figure}

To introduce the corruption in the input qubits (shown in Fig. \ref{fig:corrupt}),
the umpire uses an ancilla in state $\left|\psi\right\rangle =U(\theta,\phi,\lambda)\left|0\right\rangle =\cos\frac{\theta}{2}\left|0\right\rangle +\sin\frac{\theta}{2}\left|1\right\rangle $
prepared using single qubit unitary operation $U(\theta,\phi=0,\lambda=0)$
defined as

\begin{equation}
U(\theta,\phi,\lambda)=\left(\begin{array}{cc}
\cos\frac{\theta}{2} & -e^{i\lambda}\sin\frac{\theta}{2}\\
e^{i\phi}\sin\frac{\theta}{2} & e^{i\lambda+i\phi}\cos\frac{\theta}{2}
\end{array}\right).\label{eq:Unitary matrix}
\end{equation}
Subsequently, he uses this ancilla as control and applies CNOT gates
to the rest of the qubits which he sends to three players. The amount
of corruption, $x$ is related by $x=\sin^{2}\frac{\theta}{2}$. Therefore,
in what follows, we trace out the fourth qubit to obtain the payoffs
of noisy game.

\section{Results and discussion \label{sec:Results-and-discussion}}

We have performed the experiments for all classes and computed payoff
from the measurement outcomes in computational basis. We have also
obtained the output density matrices to obtain the fidelity between
theoretically desired and experimentally reconstructed states.

\subsection{Fidelity and quantum state tomography \label{subsec:Fidelity-and-Quantum}}

In quantum computation, fidelity is used to describe closeness between
two states as it is one of the distance based measures. Ideally, fidelity
between the experimental ($\rho^{E}$) and theoretical ($\sigma$)
density matrices, defined as $F(\rho^{E},\sigma)=Tr\sqrt{(\sqrt{\sigma}\rho^{E}\sqrt{\sigma)}}$
\cite{sisodia2018circuit}, is desired to be 1, but due to unavoidable
errors it is usually less than unity in most cases.

\begin{figure}
\begin{centering}
\subfloat[]{\begin{centering}
\includegraphics[scale=0.35]{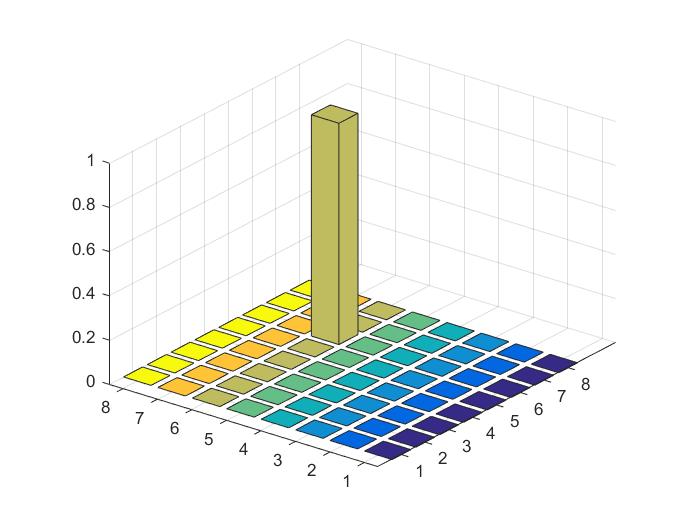}
\par\end{centering}
}\subfloat[]{\begin{centering}
\includegraphics[scale=0.35]{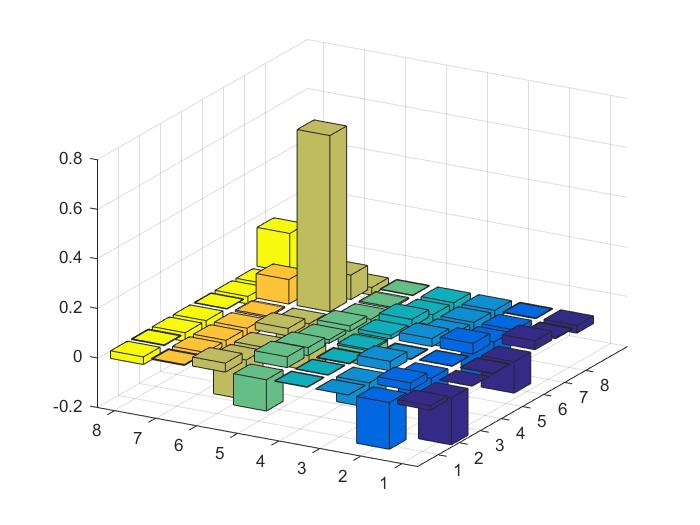}
\par\end{centering}
}
\par\end{centering}
\caption{\label{fig:QST}(Color online) Graphical representation of (a) the
theoretical and (b) the real part of the experimentally reconstructed
density matrices for Class VII.}
\end{figure}

In our case, we have calculated fidelities of all the classes and
shown them in Table 2. To obtain the experimental density matrices
and fidelities we performed quantum state tomography of the outputs
of all the circuits (see \cite{sisodia2017experimental,vishnu2018experimental}
for detail). The crux of the matter, is that we can reconstruct the three
qubit density matrix of the output of the circuit using 
\begin{equation}
\rho^{E}=\frac{1}{2^{3}}\stackrel[i_{1},i_{2},i_{3}=0]{3}{\sum}T_{i_{1}i_{2}i_{3}}(\sigma_{i_{1}}\varotimes\sigma_{i_{2}}\varotimes\sigma_{i_{3}}),\label{eq:experimental rho}
\end{equation}
where $\sigma_{j}$ are respective Pauli matrices. The values of elements
of the $T$ matrix are obtained from the expectation values of these
Pauli operators. For instance, in the case of Class VII, $U_{7}=\sigma_{x}\varotimes I\varotimes\sigma_{x}$
is the strategy unitary, the experimentally reconstructed density
matrix is shown with corresponding theoretical density matrix $\sigma=\left|101\right\rangle \left\langle 101\right|$
in Fig. \ref{fig:QST}. The fidelity of the output density matrices
for all the classes are summarized in Table \ref{tab:Results-of-experiment}. Surprisingly, fidelity and error are not properly anti-correlated, i.e., there are instances where a class of higher fidelity than another, still yields a greater error in the payoff.
Here it may be noted that the experiments are performed on different processors provided by
IBM depending upon their availability.

{\small{}{}} 
\begin{table}
\noindent \begin{centering}
{\small{}{}}%
\begin{tabular}{|c|c|c|c|c|c|c|}
\hline 
{\small{}{}Action}  & {\small{}{}Theoretical}  & {\small{}{}Experimental (IBM)}  & Experimental (NMR)  & {\small{}{}Fidelity}  & {\small{}{}Error in Payoff(\%)}  & Processor\tabularnewline
\hline 
\hline 
{\small{}{}Class I}  & {\small{}{}-3.75 }  & -2.7689  & -3.65  & 0.6753  & 26  & vigo\tabularnewline
\hline 
{\small{}{}Class II}  & {\small{}{}-3.75}  & -2.751  & -3.44  & {\small{}{}0.7813}  & 26.64  & ourense\tabularnewline
\hline 
{\small{}{}Class III}  & {\small{}{}-1.833}  & -0.7384  & -1.59  & 0.844  & 59.71  & ourense\tabularnewline
\hline 
{\small{}{}Class IV}  & {\small{}{}2}  & 2.3275  & 1.26  & 0.9236  & 16.3  & ourense\tabularnewline
\hline 
{\small{}{}Class V}  & {\small{}{}0}  & -0.0676  & 0.4  & 0.9346  & 6.76  & essex\tabularnewline
\hline 
{\small{}{}Class VI}  & {\small{}{}-5.67}  & -4.608  & -5.39  & 0.9317  & 18.73  & ourense\tabularnewline
\hline 
{\small{}{}Class VII}  & {\small{}{}6.33}  & 4.459  & 5.92  & 0.843  & 29.5  & ibmqx2\tabularnewline
\hline 
{\small{}{}Class VIII}  & {\small{}{}6.33}  & 4.03  & 6.34  & 0.8416  & 36.3  & ibm\_16\_melbourne\tabularnewline
\hline 
{\small{}{}Class IX}  & {\small{}{}4.75}  & 3.2  & 4.83  & 0.6753  & 32.6  & ibmqx2\tabularnewline
\hline 
{\small{}{}Class X}  & {\small{}{}-1.833}  & -0.8098  & -1.87  & 0.6314  & 55.8  & vigo\tabularnewline
\hline 
\end{tabular}
\par\end{centering}
\begin{centering}
 
\par\end{centering}
{\small{}{}\caption{\label{tab:Results-of-experiment} Results of the experiments performed
using different quantum processors in IBM Q Experience {for the payoff table parameters  $p=1, n=9, q=2$ considering state $x=0$. Surprisingly, fidelity and error are not properly anti-correlated, i.e., there are instances where a class of higher fidelity than another, still yields a greater error in the payoff. An example here for different processors would be Class III and Class I. Another example, for the same processor, would be, Classes IV and VI. A possible explanation could be noise during readout of payoffs. } }
}{\small\par}

\end{table}

\subsection{Payoff table \label{sec:Payoff-Table}}

Payoff of single player $\$$ is obtained by multiplying their respective
payoffs from Table \ref{tab:Generalized-Payoff-Table} with the probabilities
obtained from the output of the experiment (as in \cite{mitra2007experimental}).
The mean payoff per player $\ensuremath{\left\langle \$\right\rangle }$
is defined as the numerical mean of the payoffs of each player $\ensuremath{\left\langle \$\right\rangle }=\frac{\$1+\ensuremath{\$2+}\$3}{3}$.

It is also shown in the past that the payoffs for quantum Nash equilibrium
deteriorate with noise \cite{johnson2001playing}. However, in our
case, assuming arbitrary values of parameters we obtained that quantum
Nash equilibrium is $3\$_{{\rm qu}}=-4nx+2n+p$, and classical Nash
equilibrium for ($q$,$q$,$q$) is $\$_{{\rm cl}}=q\left(1-x\right)$.
As both quantum and classical Nash equilibrium values decrease with
corruption level $x$, the quantum advantage disappears after the
point of intersection of these two curves. That intersection can be
obtained as the critical value of corruption 
\begin{equation}
x_{c}=\frac{2n+p-3q}{4n-3q}.\label{eq:critcal corrupt}
\end{equation}

\begin{figure}
\begin{centering}
\includegraphics{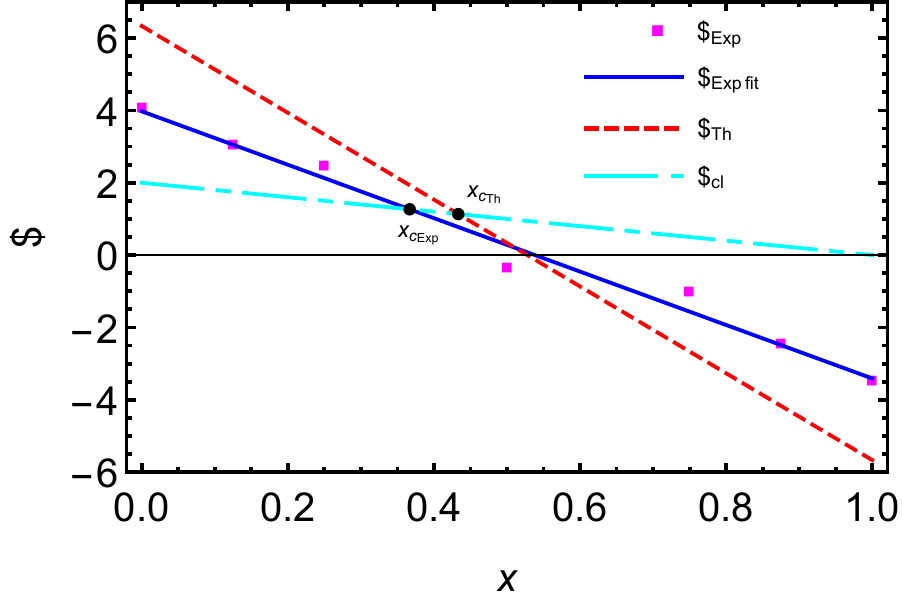} 
\par\end{centering}
{\small{}\caption{\label{fig:x-c}(Color online) Variation of payoffs in classical and
quantum Nash equilibrium with corruption $x$. We have also shown
the experimentally obtained payoff values. We assumed $p=1,\,n=9,$
and $q=2$. Here, $x_{c_{{\rm Th}}}$ and $x_{c_{{\rm Exp}}}$ represent
critical values of corruption from theoretical and experimental results. {Noise degrades quantum information, and thus not surprisingly, $x_{c_{{\rm Th}}} > x_{c_{{\rm Exp}}}$}.}
}{\small\par}

\end{figure}

From Fig. \ref{fig:x-c}, we obtained that experimentally $x_{c}=0.363$
whereas the theoretical value is $0.428$ \cite{johnson2001playing},
giving us an error of $15.18\%$. Note that the results obtained in
\cite{johnson2001playing} neglect decoherence after the initial state
is prepared by the demon. However, here on top of that, gate errors
in the implementation of the presently available SQUID based quantum
computing facilities also play an important role in sabotaging the
quantum advantage achievable in quantum games. {Of course,} a reduction
in noise with improvement in technology will improve the outcome.
Notice that for very high values of corruption, when classical strategy
is a preferred choice, the experimental results show higher payoffs
than theoretically expected in quantum Nash equilibrium. The experimental
values of payoff can be improved using mitigated error method provided
by IBM.

{\small{}{}} 

\begin{figure}
\begin{centering}
\subfloat[]{\begin{centering}
\includegraphics[scale=0.81]{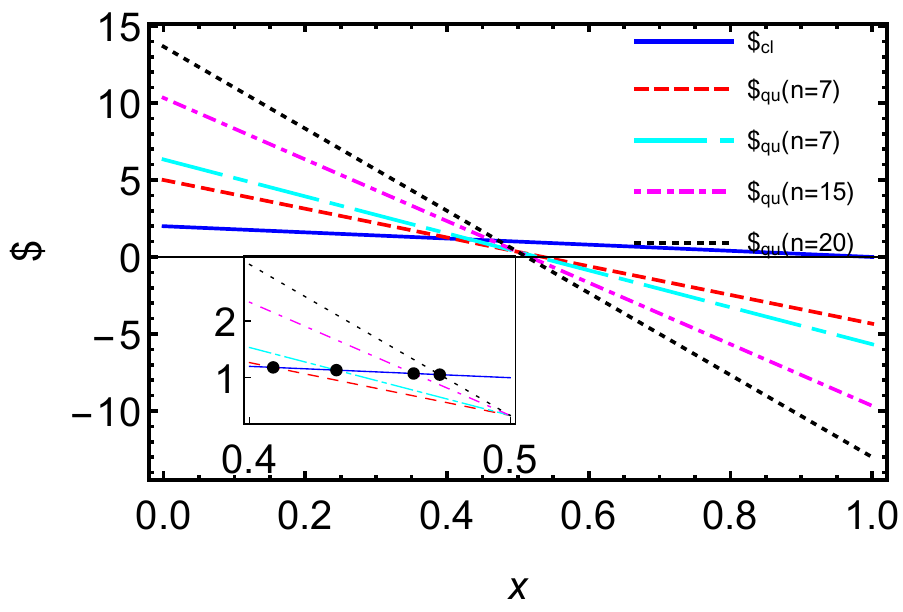} 
\par\end{centering}

}\subfloat[]{\begin{centering}
\includegraphics[scale=0.81]{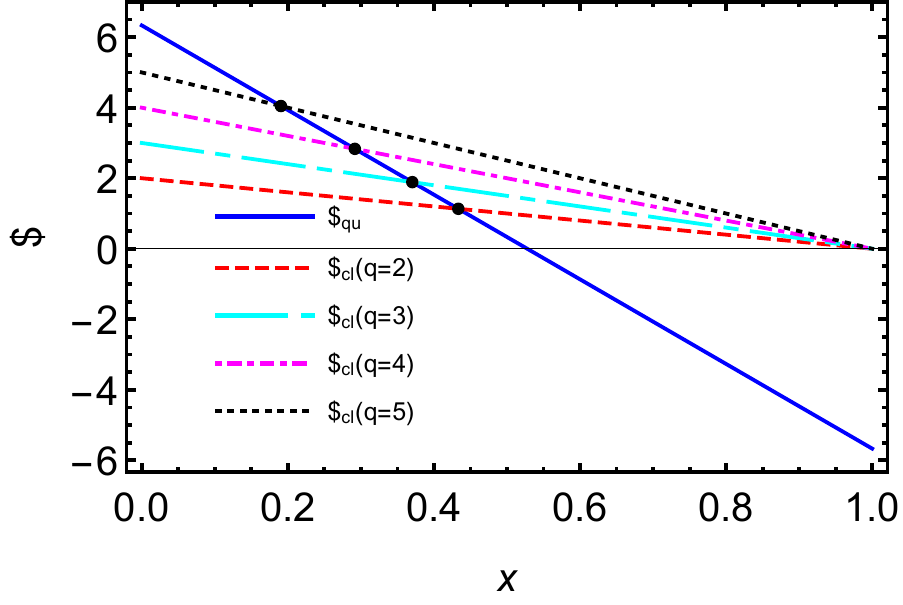}
\par\end{centering}
}
\par\end{centering}
{\small{}{}\caption{\label{fig:xc-nq}(Color online) Variation of payoffs in classical
and quantum strategies with corruption $x$ and payoff parameters
(a) $n$ ($q=2$) and (b) $q$ ($n=9$) with $p=1$ for Class VIII
considering rest of the parameters in $\left\{ p,n,q\right\} $ as
constant. The critical value of corruption $x_c$ is marked by black dots in all cases.}
}{\small\par}

\end{figure}

\subsection{Variation in $\boldsymbol{p,q,n}$ \label{subsec:Varying-}}

As we have discussed the general case of the game with arbitrary values
of the individual payoff parameters, here we discuss the role of each
of these parameters (assuming $0<p<q<n$) on $x_{c}$. {This would allow us to choose suitable payoff parameters if the noise level $x$ is known,  or if they cannot be varied, then to decide whether to employ the quantum or classical game for the problem in a practical situation.}

We find that $x_{c}$ increases (shifts to the right of the original value)
if $n$ is increased for a constant value of $p$ and $q$. It implies
that when the stakes of a game are high (large $n$), such that reward
for winnings and the amount of losses are very high, the quantum strategies
are better in spite of corruption. Further increasing $n$ saturates
 $x_{c}$ to 0.5, which signifies that no matter what, if corruption
is higher than 50\% classical strategy will always outperform the
quantum strategy. The results obtained are shown in Fig. \ref{fig:xc-nq}
(a). 

Note that the only quantum Nash equilibrium depends upon $n$, while classical
Nash equilibrium is a function of $q.$ Thus, an increase in $q$ essentially
leads to classical dominant strategy, i.e., classical strategy tends
to be as efficient as the quantum strategy (cf. in Fig. \ref{fig:xc-nq}
(b)). However, for large values of corruption there is no evident
advantage as for maximum corruption classical Nash equilibrium is
always zero.

These observation lead to conclusion that quantum systems are more
prone to errors and deteriorate rapidly with an increase in the amount
of corruption. Hence, errors in system may lead to loss of quantum
advantage originally present as observed from the experimental value
in Fig. \ref{fig:x-c} as well. Thus, in case of high errors, it is
always better to stick to classical strategies from an outsider's
perspective.

\section{Conclusion \label{sec:Conclusion-and-Discussion}}

We have discussed a multiparty quantum game by generalizing the payoff
table.  Our result may find interesting applications {in diverse fields, such as finance, social networks.}
Suppose a group of companies want to invest in a particular stock
and have limited knowledge of the market statistics, then the financial situation of the stock simulates
a three-person dilemma. In this case, if the stakes of the investments
are high such that returns are great, but so are the losses, companies
perform better if they use quantum strategies, provided the amount of preparation noise
is less than 50\%. Otherwise the classical strategy should be preferred.
Similarly, in the situation that the classical dominant strategy equilibrium
$(q,q,q)$ has relatively higher payoff than previous cases, the
present results may persuade companies to opt for classical strategies
even for a small amount of source error.

We have  performed an experiment for the noisy three-player quantum
dilemma game and observed that the obtained results were less robust against noise
than the corresponding results from NMR experiments. Further, it can be
observed that due to additional errors (other than source error introduced
in the noisy counterpart of the game) the advantage of quantum game over
corresponding classical game disappears quickly. Similar studies for
the generalizations of other games where quantum players perform better
or the games where classical strategies are always preferable can
be performed to study the role of various payoff parameters in those
cases. The present experimental implementation of the noisy quantum game
on a small noisy quantum computer establishes a {practical }quantum advantage in
game theory. However, in view of noisy intermediate-scale quantum
(NISQ) technology \cite{preskill2018quantum} around the corner, i.e.,
quantum computing infrastructure with 50-100 qubits, this advantage
can be exploited for several applications, such as in quantum machine
learning \cite{torlai2019machine}. This can be further extended to the
iterated version of the game where it is performed more than once,
and rational players decide their strategies depending upon their opponents'
previous decision. The results from the experiment performed on NMR was more
accurate, {showing that the NMR-based quantum computer is less noisy.}
To obtain a quantitative perception of that we performed
quantum state tomography here, which shows that higher fidelity of experimentally generated state does not necessarily mean smaller errors, i.e., fidelity and errors are not properly anti-correlated.

In the end, we would like to stress on the recent studies connecting
Bell nonlocality \cite{brunner2013connection,iqbal2018equivalence},
a quantum secure direct communication scheme \cite{kaur2018game},
and security of quantum key distribution schemes \cite{krawec2018game}
with game theory. In view of these works, in principle, all quantum
cryptographic schemes (see \cite{pathak2013elements,thapliyal2017quantum}
for a review) can be viewed  from the perspective of game theory, as a game to perform cryptanalysis and obtain security
proofs. For example, measurement-device-independent and device-independent as well as entangled state
based quantum key distribution schemes, such as Ekert's scheme \cite{ekert1991quantum},
can be viewed as a three-party game involving Alice, Bob and Eve.
A future work is planned to  rigorously analyze the best
strategy of Alice and Bob and that of Eve using a game theoretic approach.
We hope the present results will be helpful in the application of quantum
strategies in game theory, and in turn in their applications in
quantum technologies in general, and quantum cryptography in particular.

\section*{Acknowledgement}

AP and RS acknowledge the support from the QUEST scheme of Interdisciplinary Cyber Physical
Systems (ICPS) programme of the Department of Science and Technology
(DST), India, Grant No.: DST/ICPS/QuST/Theme-1/2019/14. KT acknowledges
the financial support from the Operational Programme Research, Development
and Education - European Regional Development Fund project no. CZ.02.1.01/0.0/0.0/16\_019/0000754
of the Ministry of Education, Youth and Sports of the Czech Republic. Authors also thank Prof.\ Anil Kumar and Dr.\ Abhishek Shukla for their interest and technical comments on this work.

\section*{Appendix: Reconstructed density matrix}

The real and imaginary parts of the experimentally obtained density matrix
by performing quantum state tomography are
\[
{\rm Re}(\rho^{E})=\left[\begin{array}{cccccccc}
0.018 & -0.188 & -0.001 & 0.002 & -0.126 & 0.036 & -0.004 & -0.031\\
-0.188 & 0.030 & -0.100 & 0.002 & 0.044 & -0.160 & -0.019 & -0.003\\
-0.001 & -0.100 & 0.037 & -0.003 & 0.033 & -0.056 & -0.049 & -0.035\\
0.002 & 0.002 & -0.003 & 0.007 & 0.015 & 0.030 & -0.042 & -0.039\\
-0.126 & 0.044 & 0.033 & 0.015 & 0.022 & -0.189 & -0.006 & -0.002\\
0.036 & -0.160 & -0.056 & 0.030 & -0.189 & 0.711 & 0.099 & -0.030\\
-0.004 & -0.019 & -0.049 & -0.042 & -0.006 & 0.099 & 0.018 & -0.022\\
-0.031 & -0.003 & -0.035 & -0.039 & -0.002 & -0.030 & -0.022 & 0.157
\end{array}\right]
\]
and 
\[
{\rm Im}(\rho^{E})=\left[\begin{array}{cccccccc}
0 & 0.229 & 0.002 & 0.009 & 0.142 & -0.042 & -0.028 & -0.009\\
-0.229 & 0 & 0 & -0.001 & -0.001 & 0.148 & -0.010 & -0.034\\
-0.002 & 0 & 0 & {0.002} & -0.011 & 0.014 & 0.053 & 0.010\\
-0.009 & 0.001 & {-0.002} & 0 & 0.014 & 0.017 & 0.004 & 0.030\\
-0.142 & 0.001 & 0.011 & -0.014 & 0 & {0.158} & 0.004 & -1.108\\
0.042 & -0.148 & -0.014 & -0.017 & {-0.158} & 0 & -1.103 & 0.058\\
0.028 & 0.010 & -0.053 & -0.004 & -0.004 & 1.103 & 0 & 0.024\\
0.009 & 0.034 & -0.010 & -0.030 & 1.108 & -0.058 & -0.024 & 0
\end{array}\right],
\]
respectively, while theoretical density matrix in the corresponding
case is given by $\sigma=\left|\psi\right\rangle \left\langle \psi\right|=\left|101\right\rangle \left\langle 101\right|$,
where $\left|\psi\right\rangle =J^{\dagger}\cdotp U_{7}\cdotp J\left|000\right\rangle $.

\begin{thebibliography}{10}
\expandafter\ifx\csname urlstyle\endcsname\relax
  \providecommand{\doi}[1]{doi:\discretionary{}{}{}#1}\else
  \providecommand{\doi}{doi:\discretionary{}{}{}\begingroup
  \urlstyle{rm}\Url}\fi
\providecommand{\bibAnnoteFile}[1]{%
  \IfFileExists{#1}{\begin{quotation}\noindent\textsc{Key:} #1\\
  \textsc{Annotation:}\ \input{#1}\end{quotation}}{}}
\providecommand{\bibAnnote}[2]{%
  \begin{quotation}\noindent\textsc{Key:} #1\\
  \textsc{Annotation:}\ #2\end{quotation}}

\bibitem{flitney2002introduction}
Flitney, A.~P., Abbott, D.: An introduction to quantum game theory. Fluctuation
  and Noise Letters \textbf{2}, R175--R187 (2002)
\input{flitney2002introduction}

\bibitem{piotrowski2003invitation}
Piotrowski, E.~W., S{\l}adkowski, J.: An invitation to quantum game theory.
  International Journal of Theoretical Physics \textbf{42}, 1089--1099 (2003)
\input{piotrowski2003invitation}

\bibitem{li2012application}
Li, X., Gao, L., Li, W.: Application of game theory based hybrid algorithm for
  multi-objective integrated process planning and scheduling. Expert Systems
  with Applications \textbf{39}, 288--297 (2012)
\input{li2012application}

\bibitem{elhenawy2015intersection}
Elhenawy, M., Elbery, A.~A., Hassan, A.~A., Rakha, H.~A.: An intersection
  game-theory-based traffic control algorithm in a connected vehicle
  environment. in: 2015 IEEE 18th international conference on intelligent
  transportation systems pp. 343--347 IEEE (2015)
\input{elhenawy2015intersection}

\bibitem{laszka2012game}
Laszka, A., Szeszl{\'e}r, D., Butty{\'a}n, L.: Game-theoretic robustness of
  many-to-one networks. in: International Conference on Game Theory for
  Networks pp. 88--98 Springer (2012)
\input{laszka2012game}

\bibitem{dodis2007cryptography}
Dodis, Y., Rabin, T., et~al.: Cryptography and game theory. Algorithmic game
  theory pp. 181--207 (2007)
\input{dodis2007cryptography}

\bibitem{li2014entanglement}
Li, A., Yong, X.: Entanglement guarantees emergence of cooperation in quantum
  prisoner's dilemma games on networks. Scientific reports \textbf{4}, 1--7
  (2014)
\input{li2014entanglement}

\bibitem{nawaz2004dilemma}
Nawaz, A., Toor, A.~H.: Dilemma and quantum battle of sexes. Journal of Physics
  A: Mathematical and General \textbf{37}, 4437 (2004)
\input{nawaz2004dilemma}

\bibitem{meyer1999quantum}
Meyer, D.~A.: Quantum strategies. Physical Review Letters \textbf{82}, 1052
  (1999)
\input{meyer1999quantum}

\bibitem{eisert1999quantum}
Eisert, J., Wilkens, M., Lewenstein, M.: Quantum games and quantum strategies.
  Physical Review Letters \textbf{83}, 3077 (1999)
\input{eisert1999quantum}

\bibitem{du2002playing}
Du, J., Xu, X., Li, H., Zhou, X., Han, R.: Playing prisoner's dilemma with
  quantum rules. Fluctuation and Noise Letters \textbf{2}, R189--R203 (2002)
\input{du2002playing}

\bibitem{werner1998optimal}
Werner, R.~F.: Optimal cloning of pure states. Physical Review A \textbf{58},
  1827 (1998)
\input{werner1998optimal}

\bibitem{tomamichel2013monogamy}
Tomamichel, M., Fehr, S., Kaniewski, J., Wehner, S.: A monogamy-of-entanglement
  game with applications to device-independent quantum cryptography. New
  Journal of Physics \textbf{15}, 103002 (2013)
\input{tomamichel2013monogamy}

\bibitem{fritz2013local}
Fritz, T., Sainz, A.~B., Augusiak, R., et~al.: Local orthogonality as a
  multipartite principle for quantum correlations. Nature communications
  \textbf{4}, 1--7 (2013)
\input{fritz2013local}

\bibitem{oppenheim2010uncertainty}
Oppenheim, J. and Wehner, S., 2010. The uncertainty principle determines the nonlocality of quantum mechanics. Science, 330(6007), pp.1072-1074.
\input{oppenheim2010uncertainty}

\bibitem{anshu2020contextuality}
Anshu, A., H{\o}yer, P., Mhalla, M., Perdrix, S.: Contextuality in multipartite
  pseudo-telepathy graph games. Journal of Computer and System Sciences
  \textbf{107}, 156--165 (2020)
\input{anshu2020contextuality}

\bibitem{popescu2014nonlocality}
Popescu, S.: Nonlocality beyond quantum mechanics. Nature Physics \textbf{10},
  264--270 (2014)
\input{popescu2014nonlocality}

\bibitem{chen2006quantum}
Chen, C., Dong, D., Dong, Y., Shi, Q.: A quantum reinforcement learning method
  for repeated game theory. in: 2006 International Conference on Computational
  Intelligence and Security volume~1 pp. 68--72 IEEE (2006)
\input{chen2006quantum}

\bibitem{clausen2018quantum}
Clausen, J., Briegel, H.~J.: Quantum machine learning with glow for episodic
  tasks and decision games. Physical Review A \textbf{97}, 022303 (2018)
\input{clausen2018quantum}

\bibitem{benjamin2001multiplayer}
Benjamin, S.~C., Hayden, P.~M.: Multiplayer quantum games. Physical Review A
  \textbf{64}, 030301 (2001)
\input{benjamin2001multiplayer}

\bibitem{johnson2001playing}
Johnson, N.~F.: Playing a quantum game with a corrupted source. Physical Review
  A \textbf{63}, 020302 (2001)
\input{johnson2001playing}

\bibitem{daskalakis2005three}
Daskalakis, C., Papadimitriou, C.~H.: Three-player games are hard. in:
  Electronic colloquium on computational complexity volume 139 pp. 81--87
  (2005)
\input{daskalakis2005three}

\bibitem{daskalakis2009complexity}
Daskalakis, C., Goldberg, P.~W., Papadimitriou, C.~H.: The complexity of
  computing a {Nash} equilibrium. SIAM Journal on Computing \textbf{39},
  195--259 (2009)
\input{daskalakis2009complexity}

\bibitem{mitra2007experimental}
Mitra, A., Sivapriya, K., Kumar, A.: Experimental implementation of a three
  qubit quantum game with corrupt source using nuclear magnetic resonance
  quantum information processor. Journal of Magnetic Resonance \textbf{187},
  306--313 (2007)
\input{mitra2007experimental}

\bibitem{kolenderski2012aharon}
Kolenderski, P., Sinha, U., Youning, L., et~al.: {Aharon-Vaidman} quantum game
  with a {Young}-type photonic qutrit. Physical Review A \textbf{86}, 012321
  (2012)
\input{kolenderski2012aharon}

\bibitem{pinheiro2013vector}
Pinheiro, A., Souza, C., Caetano, D., et~al.: Vector vortex implementation of a
  quantum game. JOSA B \textbf{30}, 3210--3214 (2013)
\input{pinheiro2013vector}

\bibitem{schmid2010experimental}
Schmid, C., Flitney, A.~P., Wieczorek, W., et~al.: Experimental implementation
  of a four-player quantum game. New Journal of Physics \textbf{12}, 063031
  (2010)
\input{schmid2010experimental}

\bibitem{zhou2003proposal}
Zhou, L., Kuang, L.-M.: Proposal for optically realizing a quantum game.
  Physics Letters A \textbf{315}, 426--430 (2003)
\input{zhou2003proposal}

\bibitem{solmeyer2018demonstration}
Solmeyer, N., Linke, N.~M., Figgatt, C., et~al.: Demonstration of a {Bayesian}
  quantum game on an ion-trap quantum computer. Quantum Science and Technology
  \textbf{3}, 045002 (2018)
\input{solmeyer2018demonstration}

\bibitem{glazer1989efficient}
Glazer, J., Ma, C.-t.~A.: Efficient allocation of a ``prize''-{King Solomon's}
  dilemma. Games and Economic Behavior \textbf{1}, 222--233 (1989)
\input{glazer1989efficient}

\bibitem{teng2013trust}
Teng, Y., Jones, R., Marusich, L., et~al.: Trust and situation awareness in a
  3-player diner's dilemma game. in: 2013 IEEE International Multi-Disciplinary
  Conference on Cognitive Methods in Situation Awareness and Decision Support
  (CogSIMA) pp. 9--15 IEEE (2013)
\input{teng2013trust}

\bibitem{bankes1994exploring}
Bankes, S.: Exploring the foundations of artificial societies: Experiments in
  evolving solutions to iterated {N}-player prisoner's dilemma. in: Artificial
  Life IV pp. 337--342 MIT Press Cambridge, MA (1994)
\input{bankes1994exploring}

\bibitem{morgan1991let}
Morgan, J.~P., Chaganty, N.~R., Dahiya, R.~C., Doviak, M.~J.: Let's make a
  deal: The player's dilemma. The American Statistician \textbf{45}, 284--287
  (1991)
\input{morgan1991let}

\bibitem{ozdemir2004quantum}
{\"O}zdemir, {\c{S}}.~K., Shimamura, J., Imoto, N.: Quantum advantage does not
  survive in the presence of a corrupt source: optimal strategies in
  simultaneous move games. Physics Letters A \textbf{325}, 104--111 (2004)
\input{ozdemir2004quantum}

\bibitem{chen2002noisy}
Chen, J.-L., Kwek, L.~C., Oh, C.~H.: Noisy quantum game. Physical Review A
  \textbf{65}, 052320 (2002)
\input{chen2002noisy}

\bibitem{chen2003quantum}
Chen, L., Ang, H., Kiang, D., Kwek, L., Lo, C.: Quantum prisoner dilemma under
  decoherence. Physics Letters A \textbf{316}, 317--323 (2003)
\input{chen2003quantum}

\bibitem{IBMQ}
{IBM Q}. {https://www.research.ibm.com/ibm-q/}. Accessed on September 2019
\input{IBMQ}

\bibitem{sisodia2017design}
Sisodia, M., Shukla, A., Thapliyal, K., Pathak, A.: Design and experimental
  realization of an optimal scheme for teleportation of an n-qubit quantum
  state. Quantum Information Processing \textbf{16}, 292 (2017)
\input{sisodia2017design}

\bibitem{sisodia2018circuit}
Sisodia, M., Shukla, A., de~Almeida, A.~A., Dueck, G.~W., Pathak, A.: Circuit
  optimization for {IBM} processors: A way to get higher fidelity and higher
  values of nonclassicality witnesses. arXiv preprint arXiv:1812.11602  (2018)
\input{sisodia2018circuit}

\bibitem{tian2015monkeys}
Tian, J., Uchida, N.: Monkeys in a prisoner's dilemma. Cell \textbf{160},
  1046--1048 (2015)
\input{tian2015monkeys}

\bibitem{shukla2018complete}
Shukla, A., Sisodia, M., Pathak, A.: Complete characterization of the directly
  implementable quantum gates used in the {IBM} quantum processors. arXiv
  preprint arXiv:1805.07185  (2018)
\input{shukla2018complete}

\bibitem{sousa2006universal}
Sousa, P.~B., Ramos, R.~V.: Universal quantum circuit for n-qubit quantum gate:
  A programmable quantum gate. arXiv preprint quant-ph/0602174  (2006)
\input{sousa2006universal}

\bibitem{sisodia2017experimental}
Sisodia, M., Shukla, A., Pathak, A.: Experimental realization of nondestructive
  discrimination of {Bell} states using a five-qubit quantum computer. Physics
  Letters A \textbf{381}, 3860--3874 (2017)
\input{sisodia2017experimental}

\bibitem{vishnu2018experimental}
Vishnu, P., Joy, D., Behera, B.~K., Panigrahi, P.~K.: Experimental
  demonstration of non-local controlled-unitary quantum gates using a
  five-qubit quantum computer. Quantum Information Processing \textbf{17}, 274
  (2018)
\input{vishnu2018experimental}

\bibitem{preskill2018quantum}
Preskill, J.: Quantum computing in the {NISQ} era and beyond. Quantum
  \textbf{2}, 79 (2018)
\input{preskill2018quantum}

\bibitem{torlai2019machine}
Torlai, G., Melko, R.: Machine-learning quantum states in the {NISQ} era.
  Annual Review of Condensed Matter Physics \textbf{11} (2019)
\input{torlai2019machine}

\bibitem{brunner2013connection}
Brunner, N., Linden, N.: Connection between {Bell} nonlocality and {Bayesian}
  game theory. Nature communications \textbf{4}, 2057 (2013)
\input{brunner2013connection}

\bibitem{iqbal2018equivalence}
Iqbal, A., Chappell, J.~M., Abbott, D.: The equivalence of {Bell's} inequality
  and the {Nash} inequality in a quantum game-theoretic setting. Physics
  Letters A \textbf{382}, 2908--2913 (2018)
\input{iqbal2018equivalence}

\bibitem{kaur2018game}
Kaur, H., Kumar, A.: Game-theoretic perspective of ping-pong protocol. Physica
  A: Statistical Mechanics and its Applications \textbf{490}, 1415--1422 (2018)
\input{kaur2018game}

\bibitem{krawec2018game}
Krawec, W.~O., Miao, F.: Game theoretic security framework for quantum key
  distribution. in: International Conference on Decision and Game Theory for
  Security pp. 38--58 Springer (2018)
\input{krawec2018game}

\bibitem{pathak2013elements}
Pathak, A.: Elements of quantum computation and quantum communication. Taylor
  \& Francis, New York (2013)
\input{pathak2013elements}

\bibitem{thapliyal2017quantum}
Thapliyal, K., Pathak, A., Banerjee, S.: Quantum cryptography over
  non-{Markovian} channels. Quantum Information Processing \textbf{16}, 115
  (2017)
\input{thapliyal2017quantum}

\bibitem{ekert1991quantum}
Ekert, A.~K.: Quantum cryptography based on {Bell's} theorem. Phys. Rev. Lett.
  \textbf{67}, 661 (1991)
\input{ekert1991quantum}

\end{thebibliography}
\end{document}